\documentclass[twocolumn,english,journal]{IEEEtran}
\usepackage[T1]{fontenc}
\usepackage[letterpaper]{geometry}
\usepackage{babel}
\usepackage{float}
\usepackage{units}
\usepackage{amsmath}
\usepackage{amsthm}
\usepackage{amssymb}
\usepackage{graphicx}
\usepackage[unicode=true,
 bookmarks=true,bookmarksnumbered=true,bookmarksopen=true,bookmarksopenlevel=1,
 breaklinks=false,pdfborder={0 0 0},pdfborderstyle={},backref=false,colorlinks=false]
 {hyperref}
\hypersetup{pdftitle={Your Title},
 pdfauthor={Your Name},
 pdfpagelayout=OneColumn, pdfnewwindow=true, pdfstartview=XYZ, plainpages=false}
\usepackage{breakurl}

\makeatletter

\floatstyle{ruled}
\newfloat{algorithm}{tbp}{loa}
\providecommand{\algorithmname}{Algorithm}
\floatname{algorithm}{\protect\algorithmname}

\theoremstyle{plain}
\newtheorem{thm}{\protect\theoremname}
\theoremstyle{plain}
\newtheorem{lem}[thm]{\protect\lemmaname}
\theoremstyle{plain}
\newtheorem{cor}[thm]{\protect\corollaryname}

\usepackage[caption=false,font=footnotesize]{subfig}

\usepackage{algpseudocode}
\usepackage{cite}

\@ifundefined{showcaptionsetup}{}{%
 \PassOptionsToPackage{caption=false}{subfig}}
\usepackage{subfig}
\makeatother

\providecommand{\corollaryname}{Corollary}
\providecommand{\lemmaname}{Lemma}
\providecommand{\theoremname}{Theorem}

\begin{document}

\title{A Lightweight, Non-intrusive Approach for Orchestrating Autonomously-managed
Network Elements }

\author{Christos Liaskos, \emph{Member, IEEE} \thanks{C. Liaskos is with the Aristotle University of Thessaloniki, Department
of Computer Science, GR-54124, AUTH Campus, Greece and with the Foundation
for Research and Technology Hellas (FORTH), Institute of Computer
Science, N. Plastira 100, Vassilika Vouton, GR-70013 Heraklion, Crete,
Greece. e-mails: cliaskos@csd.auth.gr, cliaskos@ics.forth.gr.}\vspace{-25bp}
}
\maketitle
\begin{abstract}
Software-Defined Networking enables the centralized orchestration
of data traffic within a network. However, proposed solutions require
a high degree of architectural penetration. The present study targets
the orchestration of network elements that do not wish to yield much
of their internal operations to an external controller. Backpressure
routing principles are used for deriving flow routing rules that optimally
stabilize a network, while maximizing its throughput. The elements
can then accept in full, partially or reject the proposed routing
rule-set. The proposed scheme requires minimal, relatively infrequent
interaction with a controller, limiting its imposed workload, promoting
scalability. The proposed scheme exhibits attracting network performance
gains, as demonstrated by extensive simulations and proven via mathematical
analysis.
\end{abstract}


\section{Introduction}

\IEEEPARstart{S}{oftware}-Defined Networking can imbue the network
management process with an unparalleled level of state monitoring
and control. The ability to migrate the routing elements of a network
from closed, static hardware solutions towards an open, re-programmable
paradigm is expected to promote significantly the adaptivity to demand
patterns, eventually yielding a healthy and constant innovation rate.
The OpenFlow protocol and assorted hardware \cite{McKeown.2008},
which enables an administrative authority to centrally monitor a network
and deploy fitting routing strategies, has produced significant gains
in a wide set of application scenarios \cite{Hong.2013,Agarwal.2013}.

Nonetheless, SDN-enabled traffic engineering (TE) approaches are presently
characterized by a high degree of architectural penetration. Each
networking element must yield its inner operation to a remote, central
controller. While this assumption is valid for networks managed by
the same authority (e.g. \cite{Jain.2013,Hong.2013}), it poses an
issue for networks comprising self-managed elements. Furthermore,
related solutions may come at a high capital cost, requiring multiple
powerful controllers to cover a network \cite{HassasYeganeh.2012},
as well as a high operational cost, incurred by the need for close
interaction between the network elements and the controller \cite{Tavakoli.2009},
which naturally translates to traffic overheads. These concerns, combined
with point-of-failure and security considerations \cite{McBride.2013},
can discourage self-managed elements for adopting or even trying an
SDN-based, central traffic orchestration.

The present study claims that a lightweight TE solution is in need
in order to demonstrate the gains of SDN-enabled collaboration and
gradually convince self-managed elements to participate further. The
methodology consists of applying the principles of Backpressure routing
\cite{Tassiulas.1992} to a backbone network of self-managed nodes,
deriving stability-optimal flow routing rules. Nodes that choose to
participate to the proposed scheme initially inform a central controller
of their aggregate, internal congestion states. In return, they receive
the aforementioned rule set in the form of a proposal. Apart from
its simplicity and ability to respect peering agreements, the proposed
scheme also fills a theoretical gap in the related work, offering
\emph{analytically}-proven throughput optimality and network stabilization
potential. \vspace{-10bp}

\section{Related Work\label{sec:Related-Work}}

Studies on traffic engineering in networks, whether SDN-enabled or
not, target the real-time grooming of data flows, in order to provide
the best possible quality of service on a given physical infrastructure.
To this end, maximizing the network's throughput has constituted a
prominent goal. MicroTE \cite{Benson.2011b}, Hedera \cite{AlFares.2010}
and Mahout \cite{Curtis.2011b} focus on the detection and special
handling of large \textquotedbl{}elephant\textquotedbl{} flows, under
the assumption that they constitute the usual suspects of congestion.When
a large flow is detected, it is treated as a special case, and it
is assigned a separate path, which does not conflict with the bulk
of the remaining traffic. These schemes require constant monitoring
of the network's state, which is achieved by scanning the network
for large flows via periodic polling (at the scale of $5sec$), raising
SDN controller scalability and traffic overhead concerns. They differ,
however, in where the scanning takes place. Hedera constantly scans
the edge switches of the network, requiring less nodes to visit but
more flows per node to scan. Mahout scans the hosts, scanning on average
more nodes than Hedera, but with less flows to be monitored per node.
Finally, MicroTE relies on push-based network monitoring, with nodes
posting periodically their state to the controller.

Companies have also invested in SDN-powered solutions for optimizing
their proprietary networks, within or among datacenters. Emphasis
is placed on prioritizing the applications and flows that compete
for bandwidth, based on their significance or operational requirements.
B4 \cite{Jain.2013} incorporates this concern by keeping tuples of
source, destination and QoS traits per network flow. The network's
resources are constantly monitored and the flows are assigned paths
according to their priority, breaking ties in a round-robin manner.
Microsoft's SWAN \cite{Hong.2013} considers classes of priorities,
pertaining to critical-interactive, elastic and background traffic.
Resources are first assigned per priority class. Within each coarse
assignment, a max-min fairness approach is used to distribute resources
to specific flows. Bell Labs propose a more direct approach, seeking
to solve the formal link utilization problem, given explicit flow
requests \cite{Agarwal.2013}. Other studies focus on scenarios such
as partially SDN-controlled networks, or advancing the efficiency
of multipath routing beyond classic approaches \cite{Hopps.2000},
exploiting the monitoring capabilities of OpenFlow \cite{Wojcik.2014,Domza.2015}.

Differentiating from the outlined studies, the present work proposes
a SDN-enabled traffic engineering approach that is considerably more
lightweight in terms of overhead, as well as less intrusive in terms
of architecture. Its goal is to encourage centralized, SDN-based orchestration
among autonomously managed networked elements. The proposed scheme
is throughput-optimal, yields minimal interaction with the controller
and minimal number of required flow rules.

\section{Prerequisites and System Model\label{sec:Prerequisites-and-System}}

An important term in networking studies is the notion of \emph{network
stability}. It is defined as the ability of a routing policy to keep
all network queues bounded, provided that the input load is within
the network's traffic dispatch ability, i.e. within its \emph{stability
region}. With $U_{(n,c)}(t)$ denoting the aggregate traffic accumulated
within a network node $n$ at time $t$, destined towards node $c$,
stability is formally defined as \cite[p. 24]{Georgiadis.2005}:
\begin{equation}
\underset{\tau\to+\infty}{lim\,sup}\frac{1}{\tau}\sum_{t=1}^{\tau}E\left\{ U_{(n,c)}(t)\right\} <\infty,\,\forall n,c\label{eq:stability-defintion}
\end{equation}
where $\tau$ is the time horizon and $E\left\{ *\right\} $ denotes
averaging over any probabilistic factors present in the system.

A well-developed framework for deducing network stability under a
given network management policy is the Lyapunov Drift approach. It
defines a quadratic function of the form:
\begin{equation}
L(t)=\underset{\forall n}{\sum}\underset{\forall c}{\sum}U_{(n,c)}^{2}(t)
\end{equation}
The goal is then to deduce the bounds of $\Delta L(t)=E\left\{ L(t+T)-L(t)\right\} ,$
which describes the evolution of the network queue levels over a period
$T$. The \emph{Lyapunov stability theorem} states that if it holds:
\begin{equation}
\Delta L(t)\le B-\epsilon\cdot\underset{\forall n}{\sum}\underset{\forall c}{\sum}U_{(n,c)}(t)\label{eq:LyastabilityCriteria}
\end{equation}
for two positive $B,\,\epsilon$ quantities, then the network is stable
and average queue size of inequality (\ref{eq:stability-defintion})
is bounded by $\nicefrac{B}{\epsilon}$ instead of drifting towards
infinity.

The \emph{backpressure algorithm} (BPR) defines a joint scheduling-routing
algorithm that complies with the stability criteria of inequality
(\ref{eq:LyastabilityCriteria}) and, most importantly, has been proven
to be throughput optimal \cite{Neely.2005b}. Its goal is to minimize
the lower bound of $\Delta L(t),\,\forall t$, effectively suppressing
the average queue level within the network. The analytical approach,
followed by related studies \cite{McKeown.1999}, is based on the
queue dynamics expressed by the following relation:
\begin{equation}
\underset{_{(n,c)}}{U}^{(t+T)}=max\left\{ 0,\underset{_{(n,c)}}{U}^{(t)}-O_{(n,c)}^{t\to t+T}\right\} +I{}_{(n,c)}^{t\to t+T}+G{}_{(n,c)}^{t\to t+T}\label{eq:strictQdynamics}
\end{equation}
where $O_{(n,c)}^{t\to t+T}$, $I_{(n,c)}^{t\to t+T}$ and $G_{(n,c)}^{t\to t+T}$
denotes outgoing, incoming and locally generated data at time interval
$t\to t+T$. The usual methodology then dictates a series of relaxations
of the right part of eq. (\ref{eq:strictQdynamics}), based on the
following inequalities:
\begin{equation}
O_{(n,c)}^{t\to t+T}\le\underset{l:\,source(l)=n}{\sum}\int_{t}^{t+T}\mu_{l}^{(c)}(t)\cdot dt\label{eq:relax1}
\end{equation}
\begin{equation}
I_{(n,c)}^{t\to t+T}\le\underset{l:\,destination(l)=n}{\sum}\int_{t}^{t+T}\mu_{l}^{(c)}(t)\cdot dt\label{eq:relax2}
\end{equation}
where $\mu_{l}^{(c)}(t)$ is the maximum allowed bitrate over a network
link $l$ carrying traffic destined to node $c$. Squaring both sides
of eq. (\ref{eq:strictQdynamics}) and incorporating relaxations (\ref{eq:relax1}),
(\ref{eq:relax2}), as well as the identity:
\begin{equation}
V\le max\left\{ 0,\,U-\mu\right\} +A\Rightarrow V^{2}\le U^{2}+\mu^{2}+A^{2}-2U\cdot(\mu-A)\label{eq:identity}
\end{equation}
one derives an inequality of the form of relation (\ref{eq:LyastabilityCriteria}).
Further relaxation by substituting all $\mu_{l}^{(c)}$ and $G_{(n,c)}^{t\to t+T}$
with maximum allowed values yields compliance with the Lyapunov stability
theorem. Furthermore, it is deduced that the upper bound of relation
(\ref{eq:LyastabilityCriteria}) can be minimized when maximizing
the quantity:
\begin{equation}
\underset{\forall n,k,c}{\sum}\mu_{l:source(n)\to dest(k)}^{(c)}(t)\cdot\left(U_{(n,c)}(t)-U_{(k,c)}(t)\right)\label{eq:optPursuitBPR}
\end{equation}
 The standard backbressure routing process, summarized for reference
as the SBPR Algorithm, expresses the optimization pursuit of relation
(\ref{eq:optPursuitBPR}). According to SPBR, at timeslot $t\to t+T$,
each network link $l$ must carry data towards node $c_{l}^{*}$,
such that:
\begin{equation}
\ensuremath{c_{l}^{*}\gets argmax_{c}\{U_{source(l)}^{(c)}(t)-U_{dest(l)}^{(c)}(t)\}}\label{eq:SPBR_STEP}
\end{equation}
Bidirectional links are considered as two separate unidirectional
links. Originally meant for use in wireless ad hoc networks, the BPR
process and its variants have found extensive use in packet switching
hardware and satellite systems due to their throughput optimality
trait \cite{McKeown.1999}. SBPR variants have adopted latency considerations
as well. Most prominently, authors in \cite{Ying.2011b} restrict
the node selection step (\ref{eq:SPBR_STEP}) of SBPR only within
a subset of links that offer a bounded maximum number of hops towards
the target. Other studies have shown that simply altering the queueing
discipline from FIFO to LIFO yields considerable latency gains \cite{Huang.2013b}.
Finally, it is worth noting that SBPR can be easily made TCP compatible
\cite{Seferoglu.}.

\subsection{System Model }

\begin{figure}[t]
\begin{centering}
\includegraphics[width=1\columnwidth]{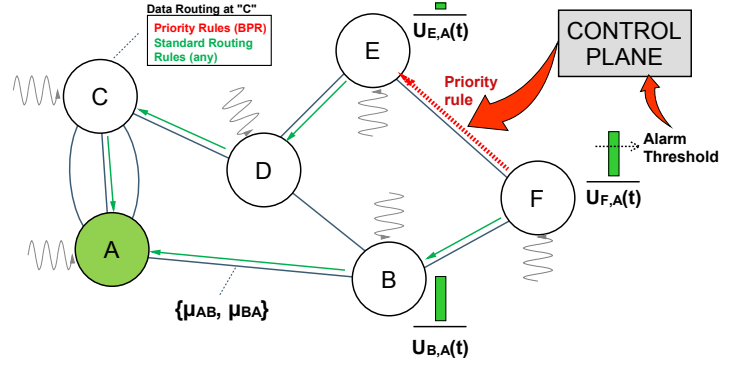}
\par\end{centering}
\caption{\label{fig:Setup}The employed system setup. A network of autonomously
managed elements, A-F, uses Backpressure-derived flow rules on top
of its standard routing scheme, in order to mitigate congestion events.
A centralized control plane orchestrates the operation of the system. }
\end{figure}

The present paper studies the use of BPR-variants in backbone networks.
The assumed setup, given in Fig. \ref{fig:Setup}, considers a network
comprising autonomously managed elements. A node can represent a single
physical router or a complete subnetwork, provided that it supports
a self-inspecting mechanism for monitoring its internal congestion
levels, as well as support a flow-based routing scheme. The nodes
are connected with links of known, time-invariant bandwidth and can
be asymmetric or unidirectional with no restriction. Due to this assumption,
the notation $\mu_{(n,c)}(t)$ is simplified to $\mu_{(n,c)}$. Data
is classified by the originating network identifier (e.g. $A$), with
no further sub-categorization.

The formed network may have any traffic-invariant traffic policy,
such as distance vector or shortest path routing. The BPR approach
operates on top of the underlying routing scheme and is enforced by
a centralized controller, which can receive node state information
and propose the installation of priority flow rules. An example is
shown in Fig. \ref{fig:Setup}. At time moment $t$, the controller
has assembled a snapshot of the network's state and notices that $U_{(F,A)}(t)$
at node $F$ exceeds a predefined alarm threshold. A BPR-variant is
executed, which deduces that traffic from $F$ towards $A$ should
better be offloaded to neighboring node $E$ for the time being. A
corresponding routing instruction is given to node $F$, which takes
precedence over all other routing rules pertaining to link $l_{FE}$.
Operation is then resumed until the next network state snapshot is
received.

OpenFlow-based solutions are most prominent candidates for the control
plane and the interaction with the network nodes \cite{McKeown.2008}.
In this case, network monitoring can be accomplished by several polling
techniques \cite{Chowdhury.2014,Yu.2013}. Without loss of generality,
we will assume that the controller obtains a consistent network state
with period $T$ \cite{Tootoonchian.2010}.

Peering agreements and routing preferences among nodes are also allowed.
For example, returning to the example of Fig. \ref{fig:Setup}, the
controller would not propose the illustrated flow rule if it was disallowed
by the peering policy/agreement between $F$ and $E$. In other words,
when the BPR-variant searches for neighbors $s\in S:\{U_{(s,A)}(t)<U_{(F,A)}(t)\}$,
the search is assumed to be limited to nodes that comply to any form
of policy, preference of agreement.

Finally, targeting minimal controller load, we allow for at most one
priority flow rule per physical network link. \vspace{-5bp}

\section{Analysis\label{sec:Analysis}}

We begin the analysis by simplifying the RHS of relation (\ref{eq:relax2}),
based on the fact that the network links have time-invariant bandwidth:
\begin{equation}
I_{(n,c)}^{t\to t+T}\le\underset{l:\,d(l)=n}{\sum}\int_{t}^{t+T}\mu_{l}^{(c)}dt=T\cdot\underset{l:\,d(l)=n}{\sum}\mu_{l}^{(c)}\label{eq:relax1-1}
\end{equation}
The RHS of relation (\ref{eq:relax1}) is simplified even further,
given that all traffic from a node $n$ towards a given destination
$c$ is served by a single outgoing link, regardless of the enforcement
of any BPR priority rules:
\begin{equation}
O_{(n,c)}^{t\to t+T}\le\underset{l:\,s(l)=n}{\sum}\int_{t}^{t+T}\mu_{l}^{(c)}dt=T\cdot\mu_{l_{nb(n)}}^{(c)}\label{eq:relax2-1}
\end{equation}
where $b(n)$ is a neighboring node of $n$ complying with any bilateral
agreements. Furthermore, applying identity (\ref{eq:identity}) to
eq. (\ref{eq:strictQdynamics}) produces:
\begin{multline}
U_{(n,c)}^{2}(t+T)\le U_{(n,c)}^{2}(t)+\left[O_{(n,c)}^{t\to t+T}\right]^{2}+\left[I_{(n,c)}^{t\to t+T}+G{}_{(n,c)}^{t\to t+T}\right]^{2}\\
-2\cdot U_{(n,c)}(t)\cdot\left[O_{(n,c)}^{t\to t+T}-I{}_{(n,c)}^{t\to t+T}-G{}_{(n,c)}^{t\to t+T}\right]
\end{multline}
Using the updated relaxations (\ref{eq:relax1-1}) and (\ref{eq:relax2-1})
and setting $\Delta U_{(n,c)}^{2}(t)=U_{(n,c)}^{2}(t+T)-U_{(n,c)}^{2}(t)$
for brevity:
\begin{multline}
\Delta U_{(n,c)}^{2}(t)\le T^{2}\cdot\left(\mu_{l_{nb(n)}}^{(c)}\right)^{2}+\left[T\cdot\underset{l:\,d(l)=n}{\sum}\mu_{l}^{(c)}+G{}_{(n,c)}^{t\to t+T}\right]^{2}\\
-2\cdot U_{(n,c)}(t)\cdot\left[T\cdot\mu_{l_{nb(n)}}^{(c)}-T\cdot\underset{l:\,d(l)=n}{\sum}\mu_{l}^{(c)}-G{}_{(n,c)}^{t\to t+T}\right]\label{eq:RHS_raw}
\end{multline}
It is not difficult to show that the RHS of inequality (\ref{eq:RHS_raw})
can be reorganized as:
\begin{multline}
\Delta U_{(n,c)}^{2}(t)\le\left[T\cdot\underset{l:\,d(l)=n}{\sum}\mu_{l}^{(c)}+U_{(n,c)}(t)+G{}_{(n,c)}^{t\to t+T}\right]^{2}\\
+\left[T\cdot\mu_{l_{nb(n)}}^{(c)}-U_{(n,c)}(t)\right]^{2}-2\cdot U_{(n,c)}^{2}(t)
\end{multline}
Summing both sides $\forall n,c$ and reminding that $\Delta L(t)=\underset{\forall n}{\sum}\underset{\forall c}{\sum}\Delta U_{(n,c)}^{2}(t)$:
\begin{multline}
\Delta L(t)\le\underset{\forall n}{\sum}\underset{\forall c}{\sum}\underset{(B)}{\left[T\cdot\underset{l:\,d(l)=n}{\sum}\mu_{l}^{(c)}+U_{(n,c)}(t)+G{}_{(n,c)}^{t\to t+T}\right]^{2}}\\
+\underset{\forall n}{\sum}\underset{\forall c}{\sum}\underset{(A)}{\left[T\cdot\mu_{l_{nb(n)}}^{(c)}-U_{(n,c)}(t)\right]^{2}}-2\cdot\underset{\forall n}{\sum}\underset{\forall c}{\sum}U_{(n,c)}^{2}(t)\label{eq:earlyInsights}
\end{multline}
We proceed by considering the RHS of relation (\ref{eq:earlyInsights})
as a function of the BPR-derived routing decisions $\mu_{l_{nb(n)}}^{(c)}$
and attempt a straightforward optimization. The $\mu_{l_{nb(n)}}^{(c)}$
can be initially treated as continuous variables. Once optimal values
have been derived, they can be mapped to the closest of the actually
available options within the network topology. The sufficient conditions
for the presence of a minimum are:
\begin{equation}
\frac{\partial RHS_{(14)}}{\partial\mu_{l_{nb(n)}}^{(c)}}=0\,\,(a),\,\,\mathcal{H}\left(\frac{\partial RHS_{(14)}}{\partial\mu_{l_{nb(n)}}^{(c)}\cdot\partial\mu_{l_{kb(k)}}^{(c)}}\right)\in\mathbb{R}^{+}\,\,(b)\label{eq:hessian}
\end{equation}
where $k$ denotes a node, $\mathcal{H}$ is the Hessian matrix \cite{HiriartUrruty.1984}
and the $0<\mathcal{H}<\infty$ refers to each of its elements. From
condition (\ref{eq:hessian}-a) we obtain:
\begin{multline}
T\cdot\left[\underset{l:\,d(l)=b(n)}{\sum}\mu_{l}^{(c)}+\mu_{l_{nb(n)}}^{(c)}\right]\\
-\left[U_{(n,c)}(t)-\left(U_{(b(n),c)}(t)+G{}_{(b(n),c)}^{t\to t+T}\right)\right]=0,\,\forall n,c\label{eq:optimization_goal}
\end{multline}
For condition (\ref{eq:hessian}-b), it is not difficult to show that
it is satisfied due to:
\begin{equation}
\frac{\partial RHS_{(\ref{eq:earlyInsights})}}{\partial\mu_{l_{nb(n)}}^{(c)}\cdot\partial\mu_{l_{kb(k)}}^{(c)}}\propto T^{2}>0,\,\forall n,k
\end{equation}
Equation (\ref{eq:optimization_goal}) represents a generalization
over the SBPR Algorithm, which operates by equation (\ref{eq:SPBR_STEP}).
At first, the eq. (\ref{eq:optimization_goal}) defines a linear system
with discrete variables $\mu_{l_{nb(n)}}^{(c)}$ and can be solved
as such. However, interesting approximations can be derived, which
also exhibit the dependence of the optimal solution from the network
topology and traffic statistics.

Firstly, the term $T\cdot\left[\underset{l:\,d(l)=b(n)}{\sum}\mu_{l}^{(c)}+\mu_{l_{nb(n)}}^{(c)}\right]$
represents the aggregate, transit traffic served by node $b(n)$,
i.e. the neighbor of $n$ that will be the recipient of traffic destined
towards $c$. A node may assume transit duties in the network, either
due to its business logic, or due to its central placement in the
topology. On the other hand, the term $\left[U_{(n,c)}(t)-\left(U_{(b(n),c)}(t)+G{}_{(b(n),c)}^{t\to t+T}\right)\right]$
refers to the role of node $b(n)$ as generator of new traffic. The
quantity $G_{(b(n),c)}^{t\to t+T}$ also introduces dependence from
traffic prediction. Indeed, at time $t$ the controller must obtain
an approximation of the traffic that will be generated at node $b(n)$
within the interval $[t,t+T]$. In other words, equation (\ref{eq:optimization_goal})
introduces a comparison between the transit and content provider aspects
of the network nodes, requiring equally balanced roles. This conclusion
is summarized in the following Lemma.
\begin{lem}
Network-wide optimization of throughput requires routing decisions
that equalize the transit and content provider roles of the nodes.
\end{lem}
Assuming a network of nodes where data transit prevails over content
generation per node, it will hold:
\begin{equation}
\underset{l:\,d(l)=b(n)}{\sum}T\left(\mu_{l}^{(c)}+\mu_{l_{nb(n)}}^{(c)}\right)>\underset{_{(n,c)}}{U}(t)-\left(\underset{_{(b(n),c)}}{U}(t)+G{}_{(b(n),c)}^{t\to t+T}\right)\label{eq:transitAssumption}
\end{equation}
for all $n,c$. In this case, the best approach for approximately
upholding equation (\ref{eq:optimization_goal}) is to maximize the
quantity
\begin{equation}
\Delta_{(n,c)}(t)=\left[U_{(n,c)}(t)-\left(U_{(b(n),c)}(t)+G{}_{(b(n),c)}^{t\to t+T}\right)\right]\label{eq:DeltaU}
\end{equation}
which depends on the traffic generated locally at node $n(b)$ during
$[t,t+T]$. In other words, the throughput-optimizing routing decision
at node $n$, regarding traffic destined to node $c$ are derived
as follows:
\begin{equation}
n^{*}=argmax_{b(n)}\left\{ \Delta_{(n,c)}(t)\right\} \label{eq:foresightOptimal}
\end{equation}
where $n^{*}$ is the optimal neighboring node of $n$ to offload
data towards $c$.

We notice that the transit assumption of (\ref{eq:transitAssumption})
is also implied by the SBPR Algorithm. Specifically, SBRP implies
that $G_{(b(n),c)}^{t\to t+T}$ is uniform for all nodes in the network,
reducing equation (\ref{eq:foresightOptimal}) to (\ref{eq:SPBR_STEP}).
This limitation is alleviated by the proposed, Foresight-enabled Backpressure
Routing (Algorithm \ref{alg:FBPR}) which targets backbone networks,
where the transit assumption of relation (\ref{eq:transitAssumption})
is expected to hold.
\begin{algorithm}[t]
\begin{algorithmic}[1]
\Procedure{FBPR}{$network\_state(t=mT|m\in\mathbf{N})$}
\For {each node $n$}
\Comment Define priority flows.
\State $visited[c]\gets 0, \forall c$
\For {each link $l : source(l)=n$}
\State $c_l^{*}(t)\gets \underset{c:!visited[c]}{argmax}\{U^{(c)}_{n}-(U^{(c)}_{d(l)}+G_{(d(l),c)}^{t\to t+T})\}$
\State $visited[c_l^{*}(t)]\gets 1$
\State $\Delta Q^{*}_l(t)\gets max\{0,U^{(c_l^{*}(t))}_{n}-U^{(c_l^{*}(t))}_{d(l)}\}$
\EndFor
\EndFor \Comment Consider multi-links, if any.
\State $\mathbf{\mu}^{*}(t)\gets argmax_{\mathbf{\mu}}\sum_{\forall l}\mu_{l}\cdot \Delta Q^{*}_{l}(t)$
\For {each link $l:\Delta Q^{*}_l(t)>0$}
\State Deploy rule $\{from:s(l),to:c_l^{*}(t),via:l\}$.
\EndFor
\State \textbf{return}
\EndProcedure
\end{algorithmic}

\caption{\label{alg:FBPR}The proposed Foresight-enabled Backpressure Routing
algorithm.}
\end{algorithm}

Line $5$ of the proposed Algorithm reflects the outcome of equation
(\ref{eq:foresightOptimal}). Inspired by \cite{Ying.2011b}, we note
that line $5$ only considers possible nodes $c$ towards which the
number of hops does not increase over link $l$. This approach favors
latency and disallows routing loops. If an alarm level is defined,
the search in line $5$ is restricted further within $c:\,U_{n}^{(c)}\ge alarm\_level$.
The $visited[.]$ array is also introduced, to make sure that each
possible destination is routed via one link at most, at each node.
The optimization of line $10$ pertains to the treatment of multi-links
that may exist in the network. Assume a triple link $M=\left\{ l_{1}:\mu_{1},\,l_{2}:\mu_{2},\,l_{3}:\mu_{3}\right\} $
and a corresponding set of $c_{l}^{*}(t)$ assignments $A=\left\{ c_{l_{1}}^{*}(t),\,c_{l_{2}}^{*}(t),\,c_{l_{3}}^{*}(t)\right\} $.
Line $10$ refers to the optimal reordering of the assignments out
of all possible $M\times A$ combinations and for each multi-link
of the network, maximizing the expected throughput. Finally, lines
$11-13$ install the FBPR-derived priority rules to the corresponding
nodes.
\begin{cor}
FBPR is throughput-optimal.
\end{cor}
We notice that the preceding analysis takes place before the relaxation
of equation (\ref{eq:optPursuitBPR}) of the classic analytical procedure.
Applying this final relaxation to equation (\ref{eq:earlyInsights})
leads to compliance with the Lyapunov stability criterion (relation
(\ref{eq:LyastabilityCriteria})) to the proof of throughput optimality,
as detailed in \cite{Neely.2005b}.

\section{Simulations\label{sec:Evaluation}}

In this Section, the performance of the proposed schemes is evaluated
in various settings, in terms of achieved average throughput, latency
and traffic losses. Specifically, the ensuing simulations, implemented
on the AnyLogic platform \cite{XJTechnologies.2013}, focus on: i)
The performance and stability gains arising from the combination of
BPR-based and Shortest Path-based (OPSF) policies, ii) The gains of
Foresight-enabled BPR over its predecessors.

The simulations assume $25$ autonomously managed nodes, arranged
in a $5\times5$ grid. Each node $n_{ij}$, $i=1\ldots5$, $j=1\ldots5$
is connected to its four immediate neighbors, $n_{i+1,j}$, $n_{i-1,j}$,
$n_{i,,j+1}$, $n_{i,,j-1}$, where applicable. This type of topology
is chosen to ensure a satisfactory degree of path diversity, i.e.
a good choice of alternative paths to connect any two given nodes.
We note that path diversity is a prerequisite for efficient traffic
engineering in general. Each link connecting two nodes is bidirectional
with $2GBps$ bandwidth at each direction.

Given that packet-level simulation of backbone networks in not easily
tractable in terms of simulation runtimes \cite{Curtis.2011b,AlFares.2010},
we assume slotted time ($1sec$ slot duration) and traffic organized
in $100MB$-long batches. At each slot, a number $G_{ij}$ of batches
is generated at each node, expressing concurrent traffic generated
from multiple internal users. The destination of each batch is chosen
at random (uniform distribution). Then, traffic batches are dispatched
according to the routing rules and the channel rates. Each node is
assumed to keep track of its internal congestion level and push it
with report/actuation period $T$ to a central controller (e.g. like
\cite{Benson.2011b}). $G_{ij}$ and $T$ are set or varied per experiment.
A node is assumed to reject/drop incoming or generated traffic when
it has more than $500$ batches on hold, using a single queue model.
Finally, the BPR schemes are enabled on a node where the number of
batches on hold exceed a certain $alarm$ $level$, set per experiment.
The $alarm\_level$ can also be perceived as a parameter that defines
whether the adoption of the BPR priority flow rules is partial or
global.

When enabled, the BPR-derived routing rules handle the enqueued batches
in a LIFO manner, as advised in \cite{Huang.2013b}. This holds for
both SBPR and FBPR in the ensuing comparisons. The Open-Shortest-Path-First
(OSPF) approach is used as the underlying DVR routing scheme in all
applicable cases. Finally, while $FBPR-OSPF$ is loop-free due to
the described, hop-based filtering at line $5$ of Algorithm \ref{alg:FBPR},
the routing rules proposed by $SBPR-OSPF$ may create loops. Therefore,
a pairwise check is performed among the nodes for the detection of
loops. If one exists, the specific BPR-derived priority routing rules
that caused it are filtered-out and are not forwarded to the nodes.

\begin{figure}[t]
\begin{centering}
\subfloat[\label{fig:T5_A20_Delay}Achieved average batch delivery times.]{\centering{}\includegraphics[width=1\columnwidth]{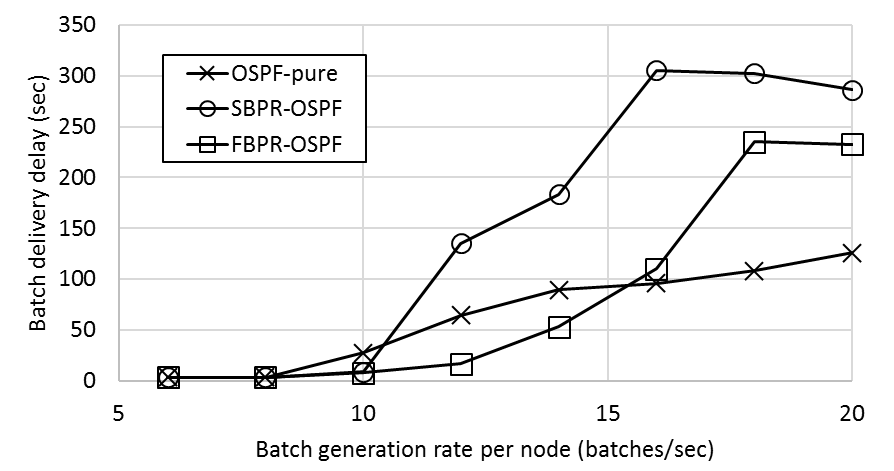}}
\par\end{centering}
\begin{centering}
\subfloat[\label{fig:T5_A20_OVF}Achieved average batch overflow rates.]{\centering{}\includegraphics[width=1\columnwidth]{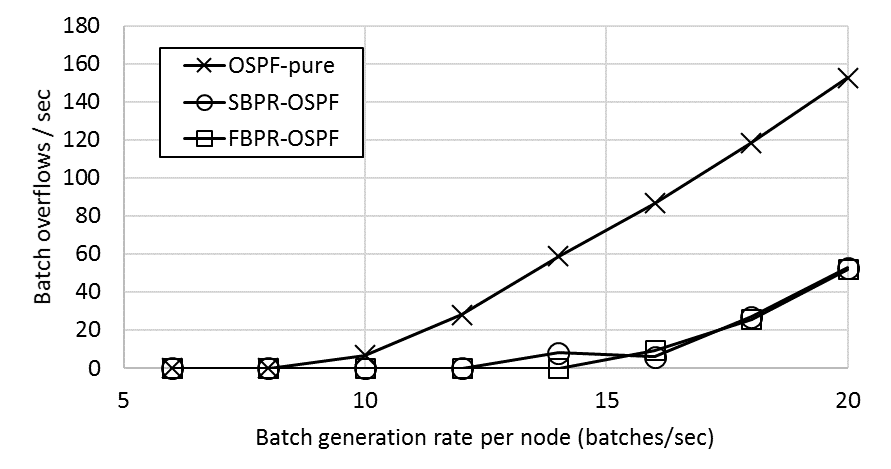}}
\par\end{centering}
\begin{centering}
\subfloat[\label{fig:T5_A20_THPT}Achieved average throughput.]{\centering{}\includegraphics[width=1\columnwidth]{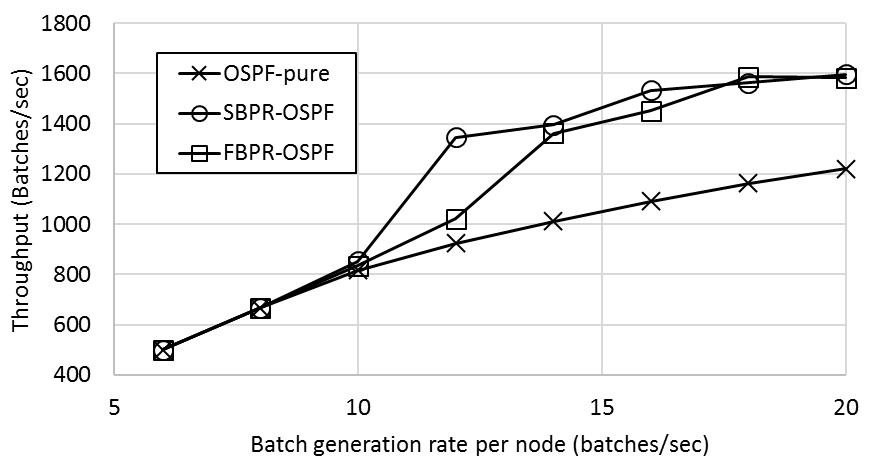}}
\par\end{centering}
\caption{\label{fig:T5_A20}The comparative performance of the Backpressure-based
schemes and a standalone OSPF approach. The alarm level is $20\%$
and actuation period $T=5sec$.}

\end{figure}

Figure \ref{fig:T5_A20} illustrates the performance of pure $OSPF$
(no overlayed BPR), $SBPR-OSPF$ and $FBPR-OSPF$, for varying network
load. The x-axis corresponds to the number of batches generated at
each node per second, $G$, which is uniform for all nodes ($G_{ij}=G,\forall i,j$).
A load of $5$ batches per second corresponds to $500MBps$ data generation
rate. For a node being serviced by four outgoing channels of $2GBps$
each, this translates to a $1:16$ channel over-subscription rate
with regard to local users only. At $G=20$ batches per second, the
ratio rises to $1:4$. The actuation period, $T$ is set to $5sec$
and the alarm level is $20\%$ of the buffer size. In terms of batch
latency, Fig. \ref{fig:T5_A20_Delay} shows that the proposed $FBPR-OSPF$
approach offers the best latency times, even over OSPF, until $G\approx17$
batches/sec. At that point, OSPF-pure offers better latency, at the
expense of an excessive traffic overflow rate (Fig. \ref{fig:T5_A20_OVF}).
As expected, dropping much of the flowing traffic benefits the delivery
times of the ``surviving'' traffic. However, all BPR-based schemes
are able to sustain operation with a limited overflow rate, even under
maximal load. In other words, the stability of the system is clearly
increased with the use of the BPR class of routing schemes. This phenomenon
is also evident from the throughput plot of Fig. \ref{fig:T5_A20_THPT}.
OSPF-pure offers the worst performance, since it leads to queue build-up
and high overflow rate. On the other hand, the proposed $FBPR-OSPF$
offers significantly improved results. Nonetheless, $SBPR-OSPF$ offers
the maximum throughput in all cases. However, given its performance
in term of latency, the superiority in raw throughput is clearly not
useful and is owed to batches traveling via excessively long routes
within the network \cite{Ying.2011b}.

\begin{figure}[t]
\begin{centering}
\subfloat[\label{fig:Complex_Delay}Average batch delivery times.]{\centering{}\includegraphics[width=1\columnwidth]{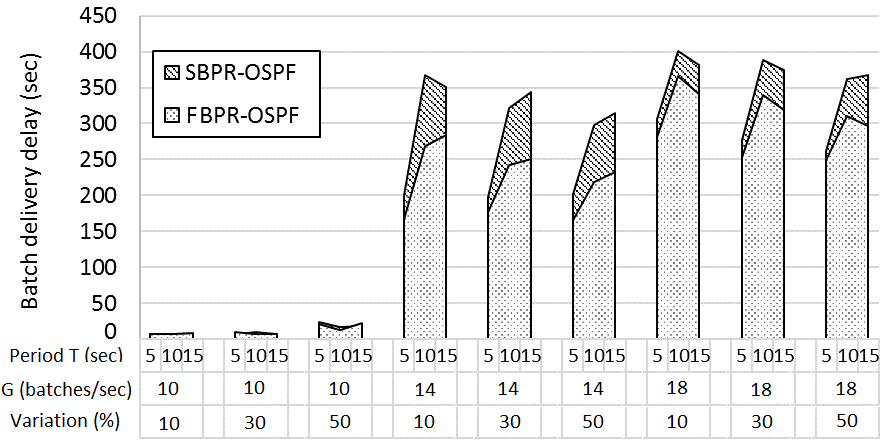}}
\par\end{centering}
\begin{centering}
\subfloat[\label{fig:Complex_OVF}Average batch overflow rates.]{\centering{}\includegraphics[width=1\columnwidth]{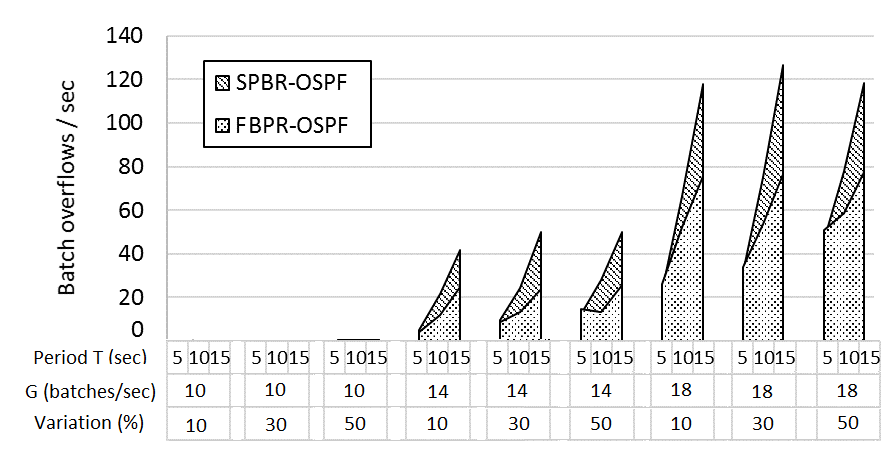}}
\par\end{centering}
\begin{centering}
\subfloat[\label{fig:Complex_THPT}Average throughput.]{\centering{}\includegraphics[width=1\columnwidth]{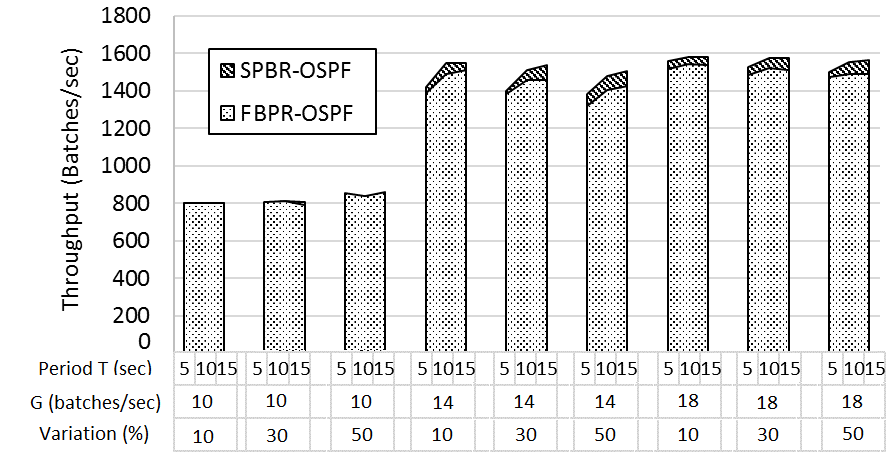}}
\par\end{centering}
\caption{\label{fig:Complex}The Foresight-enabled backpressure routing yields
significant performance gains compared to the standard backpressure
algorithm, while retaining the throughput-optimality trait.}
\end{figure}

We proceed to study the benefits of endowing BPR with foresight. In
Fig. \ref{fig:Complex}, the batch generation rate per node is set
randomly at $G_{ij}=G\pm v\cdot G$ (uniform ditribution) where $v$
is a percentage ranging from $10\%$ to $50\%$. Notice that, in the
previous experiment, FBPR and SBPR where equivalent from the aspect
of foresight, due to the constant $G$ values over all nodes. The
alarm level is kept at $20\%$ of the buffer size and $T$ is varied
from $5$ to $15$ sec. Each point in Fig. \ref{fig:Complex} is derived
as the average over $50$ simulation iterations. Since the goal of
the comparison is to deduce the gains derived from foresight, perfect
knowledge of $G_{ij}$ is passed to $FBPR-OSPF$. Furthermore, for
fairness reasons, the latency-favoring, hop-based node filtering of
FBPR is discarded. (i.e. line $5$ of Algorithm \ref{alg:FBPR} considers
all neighbors of node $n$). Thus, $FBPR-OSPF$ drops any latency
considerations that could have given an advantage over $SPBR-OSPF$
from this aspect. The performance gains in batch latency and overflow
rate are apparent in Fig. \ref{fig:Complex_Delay} and \ref{fig:Complex_OVF}
respectively. In general, the bonus of foresight is significant as
$T$ increases, since the system can make more long-lived routing
decisions. The gains are also accentuated for medium to high network
loads, where BPR in general makes sense. The trade-off between latency
and overflow rates is present in \ref{fig:Complex_Delay} and \ref{fig:Complex_OVF}
as well. Finally, the throughput optimality continues to hold (Fig.
\ref{fig:Complex_THPT}) with the slight difference being owed to
the redundant data traveling produced by SBPR. In other words, having
no foresight, SBPR takes decisions that distribute the network traffic
slightly wider, but lead to higher latency and overflow rate in the
future.

\section{Conclusion \label{sec:Conclusion-and-Future}}

The present study brought Backpressure routing (BPR) and its benefits
to the SDN-derived traffic engineering ecosystem. Its inherited benefits
include throughput maximization and optimal stability under increased
network load. The BPR and SDN combination can offer attractive, lightweight
and centrally orchestrated routing solutions. Minimum cost, non-penetrative
approaches could be the key for gradually encouraging cooperation
between distrustful autonomous parties, with significant gains for
the end-users. The presented approach can pave the way for a new class
of lightweight traffic engineering schemes that require minimal commitment
from the orchestrated network elements.

\section{Acknowledgement}

This work was funded by the EU project Net-Volution (EU338402) and
the Research Committee of the Aristotle University of Thessaloniki.

\bibliographystyle{IEEEtran}

\end{document}